\documentclass[aps,twocolumn,prb,preprintnumbers,amsmath,amssymb,superscriptaddress]{revtex4-1}

\usepackage{graphicx}
\usepackage{dcolumn}
\usepackage{bm}
\usepackage{hyperref}
\usepackage{xcolor}
\usepackage{amsmath}
\usepackage{physics}
\usepackage{mathalfa}

\usepackage{mathrsfs}
\usepackage[figuresright]{rotating}
\usepackage{amssymb}
\usepackage{dcolumn}
\usepackage{bm}
\usepackage{multirow}
\DeclareUnicodeCharacter{2212}{-}
\usepackage{natbib}
\usepackage{float}
\floatstyle{plaintop}
\restylefloat{table}
\usepackage[normalem]{ulem}

\begin{document}

\title{Quantum paramagnetism in a non-Kramers rare-earth oxide: Monoclinic $\rm Pr_2Ti_2O_7$}

\author{Huiyuan Man}
\affiliation{Institute for Quantum Matter and Department of Physics and Astronomy, The Johns Hopkins University, Baltimore, Maryland 21218, USA}

\author{Alireza Ghasemi}
\affiliation{Institute for Quantum Matter and Department of Physics and Astronomy, The Johns Hopkins University, Baltimore, Maryland 21218, USA}

\author{Moein Adnani}
\affiliation{Texas Center for Superconductivity and Department of Physics, University of Houston, Houston, Texas 77204, USA}

\author{Maxime A. Siegler}
\affiliation{Department of Chemistry, The Johns Hopkins University, Baltimore, MD 21218, USA}

\author{Elaf A. Anber}
\affiliation{Department of Materials Science and Engineering, The Johns Hopkins University, Baltimore, Maryland 21218, USA}

\author{Yufan Li}
\affiliation{Department of Physics and Astronomy, The Johns Hopkins University, Baltimore, Maryland 21218, USA}

\author{Chia-Ling Chien}
\affiliation{Department of Physics and Astronomy, The Johns Hopkins University, Baltimore, Maryland 21218, USA}
\affiliation{Institute of Physics, Academia Sinica, Taipei 11519, Taiwan}
\affiliation{Department of Physics, National Taiwan University, Taipei 10617, Taiwan}

\author{Mitra L. Taheri}
\affiliation{Department of Materials Science and Engineering, The Johns Hopkins University, Baltimore, Maryland 21218, USA}

\author{Ching-Wu Chu}
\affiliation{Texas Center for Superconductivity and Department of Physics, University of Houston, Houston, Texas 77204, USA}
\affiliation{Lawrence Berkeley National Laboratory, Berkeley, California 94720, USA}

\author{Collin L. Broholm}
\affiliation{Institute for Quantum Matter and Department of Physics and Astronomy, The Johns Hopkins University, Baltimore, MD 21218, USA}
\affiliation{Department of Materials Science and Engineering, The Johns Hopkins University, Baltimore, Maryland 21218, USA}

\author{Seyed M. Koohpayeh}
\email{koohpayeh@jhu.edu}
\affiliation{Institute for Quantum Matter and Department of Physics and Astronomy, The Johns Hopkins University, Baltimore, Maryland 21218, USA}
\affiliation{Department of Materials Science and Engineering, The Johns Hopkins University, Baltimore, Maryland 21218, USA}
\affiliation{Ralph O'Connor Sustainable Energy Institute, The Johns Hopkins University, Baltimore, Maryland 21218, USA}

\date{\today}

\begin{abstract}
Little is so far known about the magnetism of the $\rm A_2B_2O_7$ monoclinic layered perovskites that replace the spin-ice supporting pyrochlore structure for $r_A/r_B>1.78$. We show that high quality monoclinic Pr$_2$Ti$_2$O$_7$ single crystals with a three-dimensional network of non-Kramers Pr$^{3+}$ ions that interact through edge-sharing super exchange interactions, form a singlet ground state quantum paramagnet that does not undergo any magnetic phase transitions down to at least 1.8 K. The chemical phase stability, structure, and magnetic properties of the layered perovskite Pr$_2$Ti$_2$O$_7$ were investigated using x-ray diffraction, transmission electron microscopy, and magnetization measurements. Synthesis of polycrystalline samples with the nominal compositions of Pr$_2$Ti$_{2+x}$O$_7$ ($-0.16 \leq x \leq 0.16$) showed that deviations from the Pr$_2$Ti$_2$O$_7$ stoichiometry lead to secondary phases of related structures including the perovskite phase Pr$_{2/3}$TiO$_3$ and the orthorhombic phases Pr$_4$Ti$_9$O$_{24}$ and Pr$_2$TiO$_5$. No indications of site disordering (stuffing and antistuffing) or vacancy defects were observed in the Pr$_2$Ti$_2$O$_7$ majority phase. A procedure for growth of high-structural-quality  stoichiometric single crystals of Pr$_2$Ti$_2$O$_7$ by the traveling solvent floating zone method is reported. Thermomagnetic  measurements of single-crystalline Pr$_2$Ti$_2$O$_7$ reveal an isolated singlet ground state that we associate with the low symmetry crystal electric field environments that split the $(2J+1=9)$-fold degenerate spin-orbital multiplets of the four differently coordinated Pr$^{3+}$ ions into 36 isolated singlets resulting in an anisotropic temperature-independent van Vleck susceptibility at low $T$. A small isotropic Curie term is associated with 0.96(2)\% noninteracting Pr$^{4+}$ impurities.
\end{abstract}

\maketitle

\section{Introduction}

The ternary oxides of the family A$_2$B$_2$O$_7$, predominantly with the face-centered cubic pyrochlore lattice in which A$^{3+}$ is a trivalent cation and B$^{4+}$ is a transition metal, have been extensively studied due to a wide variety of interesting physical properties \cite{CrystalStructure}. These include the anomalous Hall effect in Nd$_2$Mo$_2$O$_7$ \cite{Nd2Mo2O7}, giant magnetoresistance in Tl$_2$Mn$_2$O$_7$ \cite{Tl2Mn2O7_1,Tl2Mn2O7_2,Tl2Mn2O7_3}, superconductivity in Cd$_2$Re$_2$O$_7$ \cite{Cd2Re2O7}, a phase transition from a paramagnetic metal to an antiferromagnetic insulator in Eu$_2$Ir$_2$O$_7$ \cite{Eu2Ir2O7transition}, and classic spin ice in Dy$_2$Ti$_2$O$_7$ and Ho$_2$Ti$_2$O$_7$ \cite{Balents_Nature,DTO_PP_Science,HTO_PP_Science,PZONC}. Pyrochlore magnets are also promising systems for realizing three-dimensional (3D) quantum spin liquids \cite{PZONC,PHO_neutron} and other exotic states of matter \cite{MonopoleSupersolid,PIO_Nature,PIO_Luttinger,PhysRevX.1.021002,Yb2Ti2O7_PhaseDiagram,Yb2Ti2O7_MuliMagnetism,Yb2Ti2O7_thermalHall,Yb2Ti2O7_111PhaseDiagram,PZO_WJJPRL,CationRatio,Ho2Ti2O7_YishuWang,Tb2Ti2O7_XinshuZhang,Yb2Ti2O7_Scheie}.

Among the extensively studied rare-earth titanates [RE$_2$Ti$_2$O$_7$], those which adopt the layered perovskite monoclinic structure have been relatively less explored. The formation and stability of this layered structure depends on the ratio $r\textsubscript{A}/r\textsubscript{B}$ between the radii of the A$^{3+}$ and B$^{4+}$ cations (here Pr$^{3+}$ and Ti$^{4+}$). For $r\textsubscript{A}/r\textsubscript{B}> 1.78$, which occurs for La, Ce, Pr, and Nd, the low-symmetry layered-perovskite monoclinic structure (space-group \textit{P}2$_1$) is preferred, whereas in the range of $1.48 \leq r\textsubscript{A}/r\textsubscript{B} \leq 1.78$ (from Sm to Lu and Y), the titanates form the cubic pyrochlore structure (space group \textit{Fd}$\overline{3}$\textit{m}) \cite{Lanthanide_Titanates,PTO_RamanXray}.

\begin{figure}
\centering
\includegraphics[width=8.2cm]{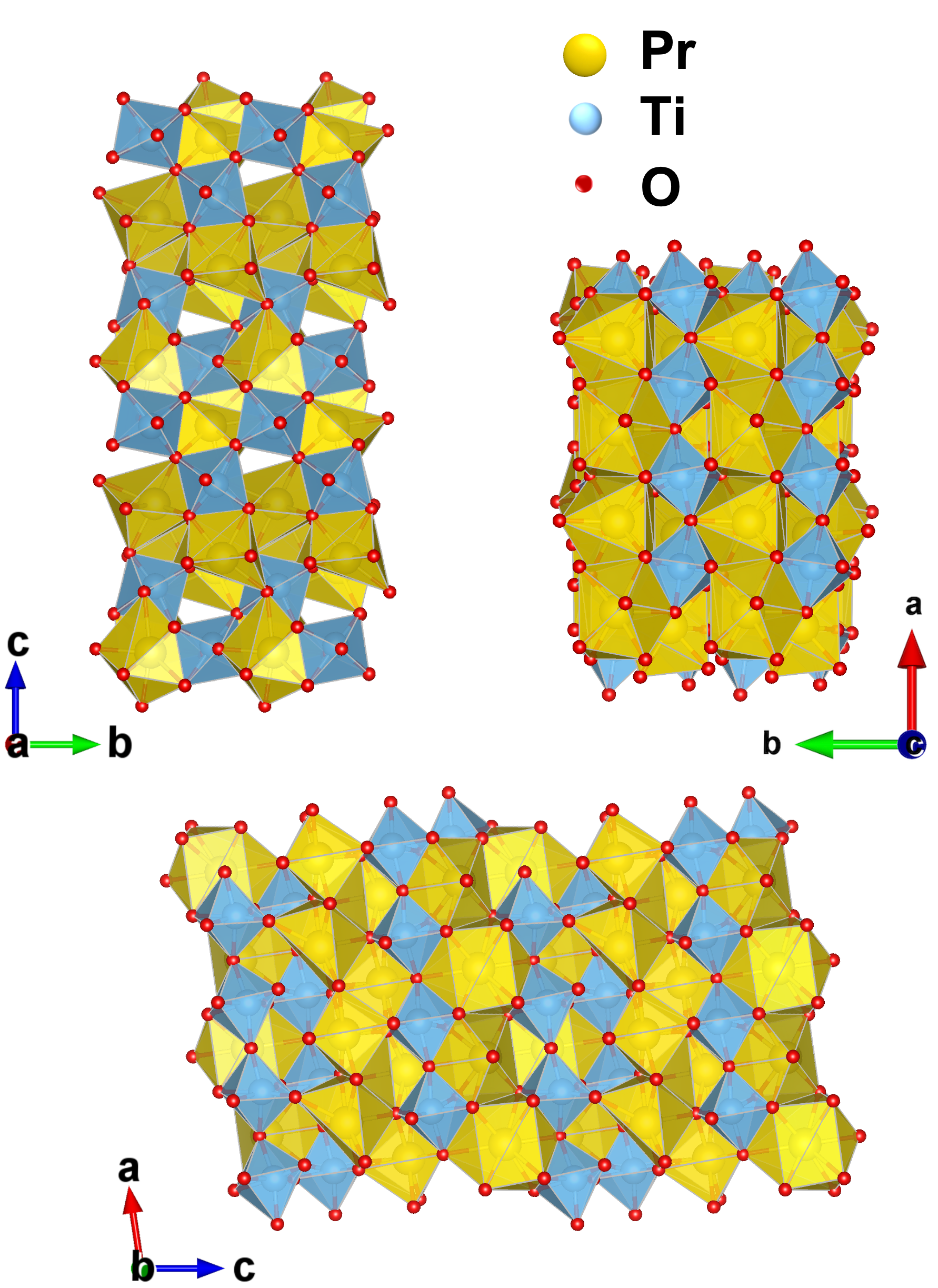}
\caption{Layered-perovskite monoclinic crystal structure of Pr$_2$Ti$_2$O$_7$. The structure consists of two-dimensional infinite perovskite slabs (four layers of corner-sharing TiO$_6$ octahedra) that are separated from each other by two praseodymium-rich layers. There are four formula units per monoclinic unit cell.}
\label{structure}
\end{figure}

The monoclinic structure is described by four (001) layers ($n=4$) of corner-sharing TiO$_6$ octahedra that form a perovskite like slab separated by two (001) layers of RE-site cations (Fig. \ref{structure}). Praseodymium occupies four $2a$ Wyckoff sites that fall in two groups.  Located in the perovskite slab within interstices defined by the TiO$_6$ octahedra, Pr1 and Pr4 are twelve-fold and seven-fold coordinated by oxygen, respectively. Pr2 and Pr3 with oxygen coordination number 10, bracket the perovskite slabs \cite{Perovskites}. Since these four Pr sites carry the low-$C_2^2$ point-group symmetry, the $J=4$ spin-orbital multiplet of Pr$^{3+}$ must be split into nine nonmagnetic singlet levels. In addition, as the four sites are different, a total of $4\times 36$ singlet-crystal-field levels should be anticipated. If the energy scale for interactions is less than the splitting between the singlet ground state and the first excited singlet, then we can expect  $\rm Pr_2Ti_2O_7$ to form a paramagnetic band insulator, which although magnetizable, should not, in general, be expected to undergo a magnetic phase transition. It should, however, be possible to drive the material to quantum criticality and an ordered state through the application of pressure or field at low temperatures.

The monoclinic layered perovskite titanates were previously studied for their interesting ferroelectric \cite{PTO_nano1,PTO_thinfilm,PTOstructure1,PTO_electronicstructure,PTO_Ferroelectricity,LTO_Ferroelectricity,LTONTO_Ferroelectricity,R2Ti2O7_ferroelectricity}, piezoelectric \cite{PTO_thinfilm}, nonlinear optical \cite{A2B2O7_single}, photocatalytic \cite{PTO_Photocatalytic,PTO_firstprinciple}, and dielectric \cite{PTO_nano1,PTO_nano2,PTO_firstprinciple} properties as well as their high ferroelectric Curie temperature\cite{PTO_Ferroelectricity,LTO_Ferroelectricity,LTONTO_Ferroelectricity}. Monoclinic Pr$_2$Ti$_2$O$_7$, in particular, was grown and studied as powder through solid-state reaction \cite{PTO_RamanXray}, nanoparticles through the sol-gel method \cite{PTO_nano1} or the modified self-propagated high-temperature synthesis method \cite{PTO_nano2}, and as epitaxial thin films grown by pulsed laser deposition \cite{PTO_thinfilm}. Single-crystal synthesis was also reported \cite{A2B2O7_single,PTOPSO_Raman}. The powder samples of Pr$_2$Ti$_2$O$_7$ showed a monoclinic structure in the noncentrosymmetric $P$2\textsubscript{1} space group and were investigated by x-ray diffraction and Raman spectroscopy \cite{PTO_RamanXray,PTOstructure1,PTOstructure2}. The electronic structure of Pr$_2$Ti$_2$O$_7$ has been studied with x-ray photoelectron spectroscopy \cite{PTO_electronicstructure,PTO_Photocatalytic}, and the optical properties were investigated by first-principles density functional theory calculations \cite{PTO_firstprinciple}. The photocatalytic activity of RE$_2$Ti$_2$O$_7$ (RE = La, Pr, and Nd) is highly dependent on their electronic band structure \cite{PTO_Photocatalytic}. Ferroelectric measurements showed a Curie temperature beyond 1555(5)~$^\circ$C\cite{PTO_Ferroelectricity}, meanwhil, similar properties were observed for monoclinic La$_2$Ti$_2$O$_7$ and Nd$_2$Ti$_2$O$_7$ with Curie temperatures of 1461(5) $^\circ$C\cite{LTO_Ferroelectricity,LTONTO_Ferroelectricity} and 1482(5) $^\circ$C\cite{LTONTO_Ferroelectricity}, respectively. The temperature-dependent Raman spectroscopy is very different for the pyrochlore “dynamic spin-ice” compound Pr$_2$Sn$_2$O$_7$ and its non-pyrochlore (monoclinic) counterpart Pr$_2$Ti$_2$O$_7$ \cite{PTOPSO_Raman}.

We conduct a comprehensive study of the phases, structures and disorders associated with the compositional deviations from the stoichiometric Pr$_2$Ti$_2$O$_7$ here. As we will see, this reveals a distinct contrast in chemical solubility compared to pyrochlore titanates. An investigation of the low-symmetry lattice structure's effect on the low-temperature magnetic properties of Pr$_2$Ti$_2$O$_7$ is lacking. In this paper, we report a systematic investigation of the phase and structural stability of Pr$_2$Ti$_2$O$_7$ based on synthesis of polycrystalline powders and single crystals. A process to develop a stoichiometric high-quality  Pr$_2$Ti$_2$O$_7$ single crystal by the traveling solvent floating zone (TSFZ) technique is reported. In order to understand the role which the lattice structure plays in determining the magnetic properties, single-crystalline Pr$_2$Ti$_2$O$_7$ is characterized through magnetization measurements versus temperature, field, and crystalline directions. Our paper shows there is no magnetic ordering and the non-Kramers Pr$^{3+}$ ions form singlet ground states with an energy gap to excited states. The presence of a three dimensional network of edge-sharing praseodymium oxide polyhedra suggest that it may be possible to induce a quantum phase transition to an ordered state through the application of pressure, strain, or high magnetic fields.

\section{Experimental details}

\subsection{Synthesis of powders}

Powder samples of praseodymium titanates with nominal compositions Pr$_2$Ti$_{2+x}$O$_7$, $-0.16\leq x\leq0.16$ were synthesized by the solid-state reaction method. The  Pr$_6$O$_{11}$ (99.99\% Alfa Aesar) and TiO$_2$ (99.99\% Alfa Aesar) starting materials were dried at 1050 $^{\circ}$C for 10 h. Pr$_6$O$_{11}$ powder was reduced to Pr$_2$O$_3$ in a hydrogen atmosphere. The resulting powders were then weighed, and the appropriate ratio was mixed, thoroughly ground, and heated to 1300 $^{\circ}$C in air for 10 h. The process of mixing, grinding, and heating was repeated four times. 

\subsection{Single-crystal growth}

Crystal growth of praseodymium titanates was carried out using the floating-zone melting technique \cite{FZ_Seyed,TSFZ_Seyed} in a four-mirror image furnace equipped with four 1 kW halogen lamps (Crystal Systems, Inc., FZ-T-4000-H-VII-VPO-PC). Powder samples of stoichiometric Pr$_2$Ti$_2$O$_7$ were first pressed into rods typically 5 mm in diameter and 8 cm in length and sintered in air at 1300 $^{\circ}$C for 10 h. The feed rod and the seed rod were mounted on the upper and lower shafts of the furnace, respectively, with the mirror stage moving upward during the growth process. The rotation speeds of the upper and lower shafts were 3 and 6 rpm, respectively, during growth.

\subsection{Characterization}

The resulting phases, crystal structures, and lattice parameters were determined using powder x-ray diffraction (XRD) at room temperature. Data were acquired for 4 covering a scattering angle range of $5^{\circ}< 2\theta <100^{\circ}$  on a Bruker D8 focus diffractometer with monochromatic Cu K$_{\alpha}$ radiation and a LynxEye strip detector. Standard Si (silicon) powder was added as a reference into the samples. Rietveld refinement of the XRD patterns was carried out using TOPAS software (Bruker AXS). 

All reflection intensities of single-crystalline Pr$_2$Ti$_2$O$_7$ were measured at 213(2) K using a SuperNova diffractometer (equipped with an Atlas detector) with Mo $K_{\alpha}$ radiation ($\lambda = 0.71073$ \AA) under the program CrysAlisPro (Version CrysAlisPro 1.171.39.29c, Rigaku OD, 2017). The same program was used to refine the cell dimensions and for data reduction. The structure was solved with the program SHELXS-2018/3 and was refined against the squared structure factor with SHELXL-2018/3\cite{Refinement_Max}. Empirical absorption correction using spherical harmonics was applied through CrysAlisPro. The sample temperature during data acquisition was controlled using a Cryojet system (manufactured by Oxford Instruments). 

The single crystals were oriented using a tungsten anode back-reflection x-ray Laue diffractometer with a 1-mm diameter x-ray beam spot. A single crystal oriented with (100), (010), and (001) planes was cut using a diamond wire saw to the dimension of $1.28\times1.58\times2.47$ mm$^3$, respectively, for a mass of 29.8 mg. For transmission electron microscopy (TEM), samples were prepared by crushing as-grown crystals in ethanol and placing a single drop of the solvent on a TEM carbon grid. We used a JEOL 2100 field emission TEM equipped with a high-resolution pole piece. Temperature-dependent susceptibility and isothermal magnetization measurements were performed using a Quantum Design superconducting quantum interference device magnetometer on the single crystal mentioned above in three different orientations and on a stoichiometric polycrystalline sample.

\section{Experimental Results}

\subsection{Polycrystalline Pr$_2$Ti$_2$O$_7$}

Contrary to the high-temperature dynamic melt/liquid-zoning crystal growth processes,  solid-state powder synthesis at lower temperatures is a highly controllable process to explore both  stoichiometric Pr$_2$Ti$_2$O$_7$ and neighboring off-stoichiometric phase(s)/structures. Detailed in the Supplementary Material \cite{SM} (see also references \cite{LTO_PhaseDiagram1,LTO_PhaseDiagram2,Parson,HistorySHELX,Pr6O11_TiO2,La_Ti_O,PZO_seyed,HTO_Ali,ETO_Wang,La2_3TiO3_x,Ln2_3TiO3,LTO_HighPressure,Ln2TiO5,Ln2TiO5_CrystalChemistry,La2_3TiO3,YTO_Seyed,AdMat,Superionics,Ln2Ti2O7_Reduction,Rigaku,SHELXSHELXL} therein), our study of the effect of variations in the net reagent composition in Pr$_2$Ti$_{2+x}$O$_7$ for ($-0.16\leq x\leq0.16$) shows the monoclinic phase forms as a stoichiometric compound mixed with secondary phases to accommodate the reagent composition. This sets Pr$_2$Ti$_{2}$O$_7$ apart from the cubic rare earth pyrochlores, which can deviate considerably from the ideal 2:2:7 stoichiometry. The stability of Pr$_2$Ti$_{2}$O$_7$ as a stoichiometric solid, facilitates the process of growing high quality stoichiometric single crystals.

\subsection{Single crystal of Pr$_2$Ti$_2$O$_7$}

\begin{figure}
\centering
\includegraphics[width=8.7cm]{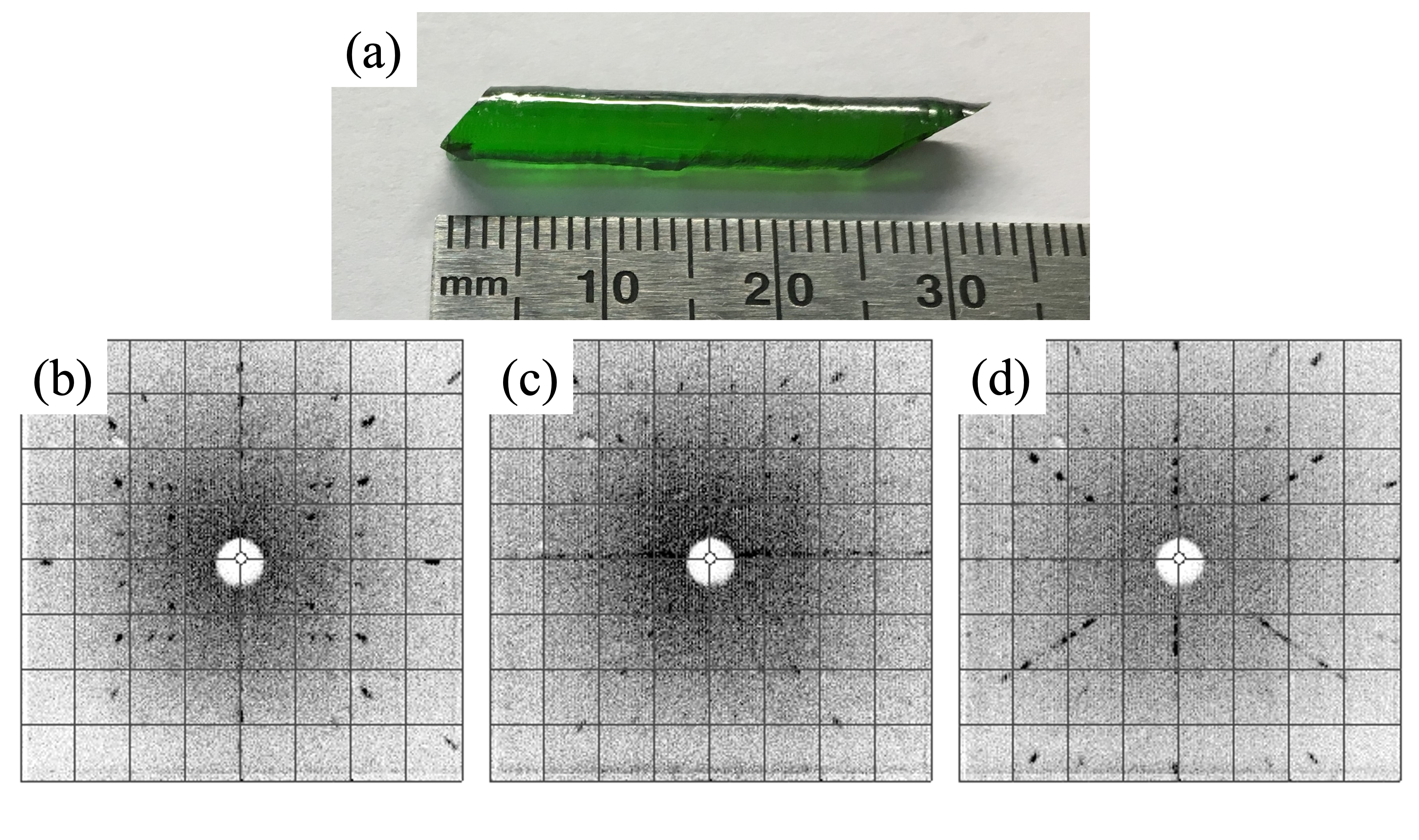}
\caption{(a) Image of the $\rm Pr_2Ti_2O_7$ single crystal grown by the TSFZ technique. (b)-(d) Laue back reflection x-ray diffraction patterns taken with the beam incident normal to the (100), (010), and (001) planes, respectively.}
\label{crystal}
\end{figure}

\begin{figure}
\centering
\includegraphics[width=8cm]{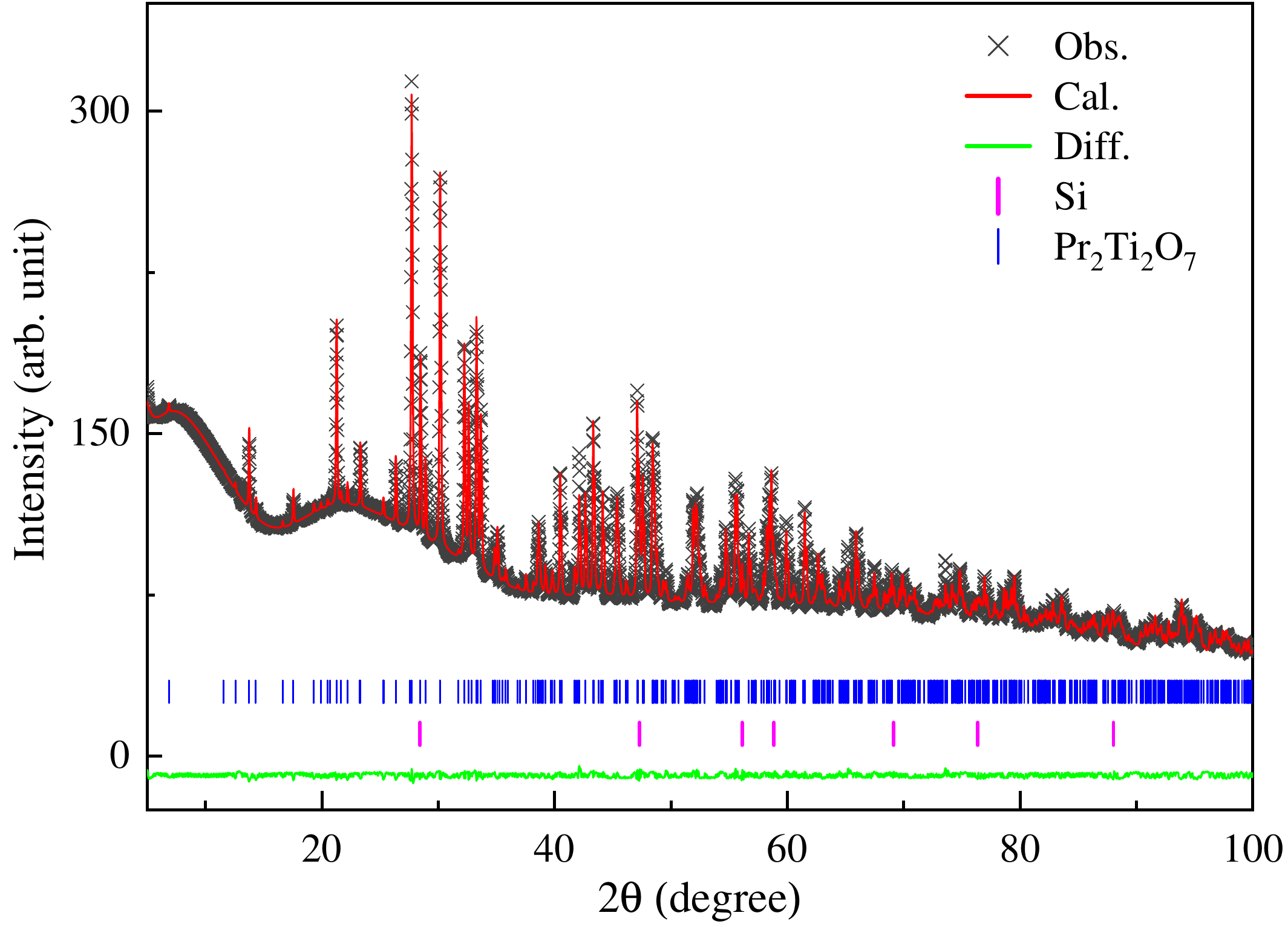}
\caption{Rietveld refinement of the XRD pattern taken from powder formed by crushing a Pr$_2$Ti$_2$O$_7$ single crystal. The collected pattern is in black, the simulated pattern is in red, and the difference is in green. Locations of Bragg reflections for Si and Pr$_2$Ti$_2$O$_7$ are marked as pink and blue vertical bars, respectively. The background arises from the quartz sample plate.}
\label{refineAli}
\end{figure}

\begin{figure}
\centering
\includegraphics[width=8cm]{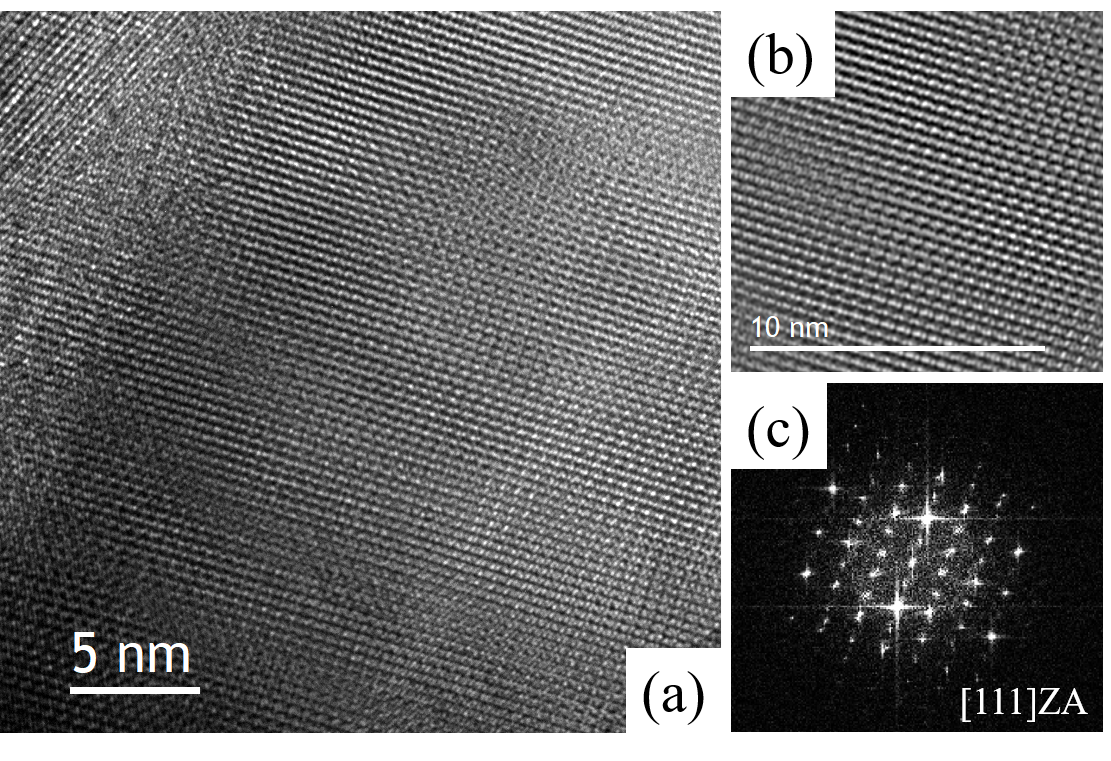}
\caption{(a) HRTEM image of the Pr$_{2}$Ti$_{2}$O$_{7}$ crystal. (b) The corresponding inverse fast Fourier transform image. (c) Fast Fourier transformation image, showing the reflection with beam incident along the [111] direction.}
\label{TEM}
\end{figure}

Although high-temperature phase instabilities and incongruent melting were encountered during the FZ melting process, crystal growths of Pr$_2$Ti$_2$O$_7$ were successfully performed by the TSFZ technique under 1-bar static ultrahigh purity argon. The growth was stable during the entire process at the consistent lamp power of 56.5\%, and no vaporization was observed. A typical transparent and green  Pr$_2$Ti$_2$O$_7$ crystal together with the Laue patterns along the main crystallographic planes are shown in Fig. \ref{crystal}. X-ray Laue patterns taken at regular intervals along the lengths and cross sections of the crystals indicated high crystalline quality with no detectable variation of orientation and no evidence of spot splitting or distortion. The crystal showed an easy cleavage along the (001) plane.

The Rietveld refinement of the XRD pattern (Fig. \ref{refineAli}) taken from the crushed crystal at room temperature confirms that the crystal is single phase, the structure is monoclinic with space group $P$2\textsubscript{1}, and the lattice constants were measured to be $a = 7.7116$, $b = 5.4867$, $c = 13.0072$ \AA, and $\beta = 98.5563^{\circ}$, similar to the stoichiometric powder sample. 

The crystal data and structure refinement for Pr$_2$Ti$_2$O$_7$ measured at 213(2) K is shown in Table S2 of the Supplemental Material \cite{SM}. The occupancy factors for PrX and TiX (X = 1-4) were all refined freely in Table S3 of the Supplemental Material \cite{SM}, and their final values are as follows: Pr1 0.970(5), Pr2 0.969(5), Pr3 0.959(5), Pr4 0.961(5), Ti1 0.966(6), Ti2 0.964(6), Ti3 0.963(6), and Ti4 0.955(6). The occupancy factors for all O atoms refine to 1 within their standard uncertainties, and were all constrained to be 1 in the final refinement.  The absolute configuration has been established by anomalous dispersion effects in diffraction measurements on the crystal, and the Flack and Hooft parameters refine to -0.024(11) and -0.026(9), respectively.

Single crystals of Pr$_2$Ti$_2$O$_7$ were further examined by high-resolution transmission electron microscopy (HRTEM) from the [111] zone axis, indicating a crystal of high structural quality (Fig. \ref{TEM}).

\subsection{Magnetization measurements}

\begin{table*}
\normalsize

\begin{center}
\renewcommand{\multirowsetup}{\centering}
{
\begin{tabular}{|c|c|c|c|c|c|c|}
\hline
Field direction ($\alpha$) & $C_{0}^{\alpha\alpha}$ ($\frac{\rm emu\ K}{\rm Oe \ mole\,Pr}$)  & $C_1^{\alpha\alpha}$ ($\frac{\rm emu\ K}{\rm Oe \ mole\,Pr}$) & $\Delta_1^\alpha$ (meV)& $C_2^{\alpha\alpha}$ ($\frac{\rm emu\ K}{\rm Oe \ mole\,Pr}$) & $\Delta_2^\alpha$ (meV)\\
\hline
$\alpha=(100)$ & 0.0054(2) & 0.50(2) & 3.93(7) & 1.03(2) & 9.2(1) \\
\hline
$\alpha=(010)$ & 0.0046(1) & 0.789(3) &	6.99(2) & 0.84(1) & 50(1)\\
\hline
$\alpha=(001)$ & 0.0050(2)  & 0.324(3) & 4.72(4) & 0.805(5) & 28.4(4) \\
\hline
Average  & 0.0050(2) &0.54(2) &5.2(1) &0.89(2) & 29.2(4)\\
\hline
Powder sample & 0.0086(3)  & 0.711(9) & 5.55(5) & 0.918(8) & 28.3(7) \\
\hline
\end{tabular}
}

\end{center}
\caption{The fitting parameters from analysis of the $T-$dependent susceptibility of monoclinic $\rm Pr_2Ti_2O_7$ for various directions ($\alpha$) of a single crystal, the average of parameters for those three directions, and the parameters obtained by fitting data for a polycrystalline sample.}
\label{fitpam}
\end{table*}

\begin{figure*}
\centering
\includegraphics[width=15cm]{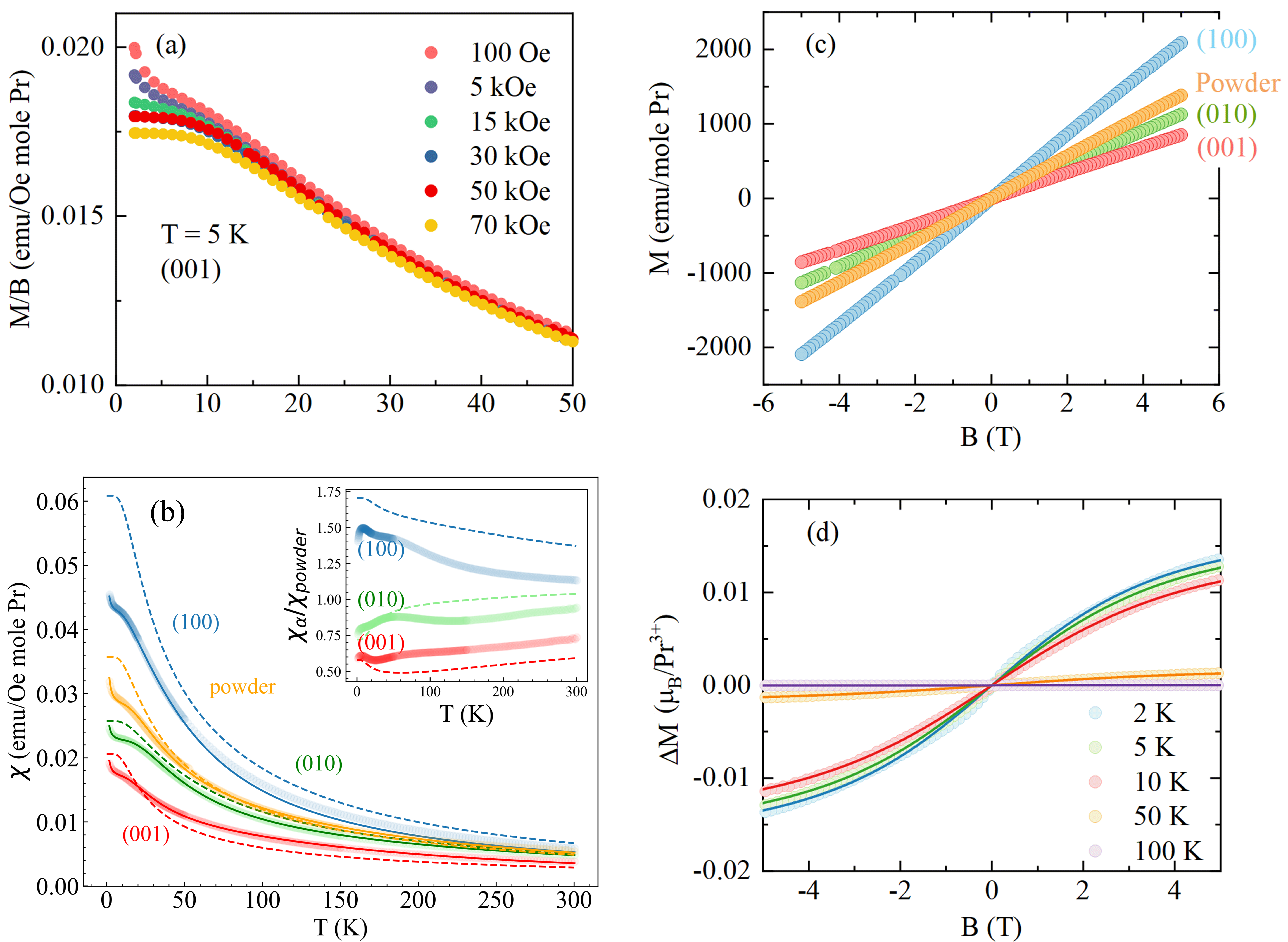}
\caption{Magnetization data for single crystal and polycrystalline samples of monoclinic $\rm Pr_2Ti_2O_7$. (a)  Temperature-dependent ``susceptibility'' $M(B,T)/B$ for fields normal to the (001) plane under the zero-field-cooled condition with multiple measurement fields. Demagnetization corrections are negligible. (b) Susceptibility inferred from magnetization measurements in a field of 1 kOe while warming from 1.8 to 300 K. The solid lines show best fits to Eq.~(\ref{hyper}) with the parameters listed in Table~(\ref{fitpam}). Dashed lines represent the susceptibility calculated from the point charged based crystal-field model as detailed in the Supplemental Material \cite{SM} and \cite{AllenCF}. The inset shows the ratio of the single crystal to the powder sample magnetic susceptibility versus $T$.  (c) Isothermal magnetization measured at 5 K for the three directions of a single crystal and for a polycrystalline sample of Pr$_{2}$Ti$_{2}$O$_{7}$. (d) The non-linear part of the isothermal magnetization $\Delta M(B,T)$ for fields normal to the (001) plane at various temperatures. The linear part determined by fitting data for $|B|>6.5$~T in (c) was subtracted. The fits to $\Delta M(B,T)$ are based on Eq.~(\ref{langevin}) and yield an impurity concentration of $f=0.96(2)\%$ with an average saturation moment of $\mu_{sat}=2.00(3)~\mu_B$.}
\label{SUSCfit}
\end{figure*}
\begin{figure}
\includegraphics[width=8.7cm]{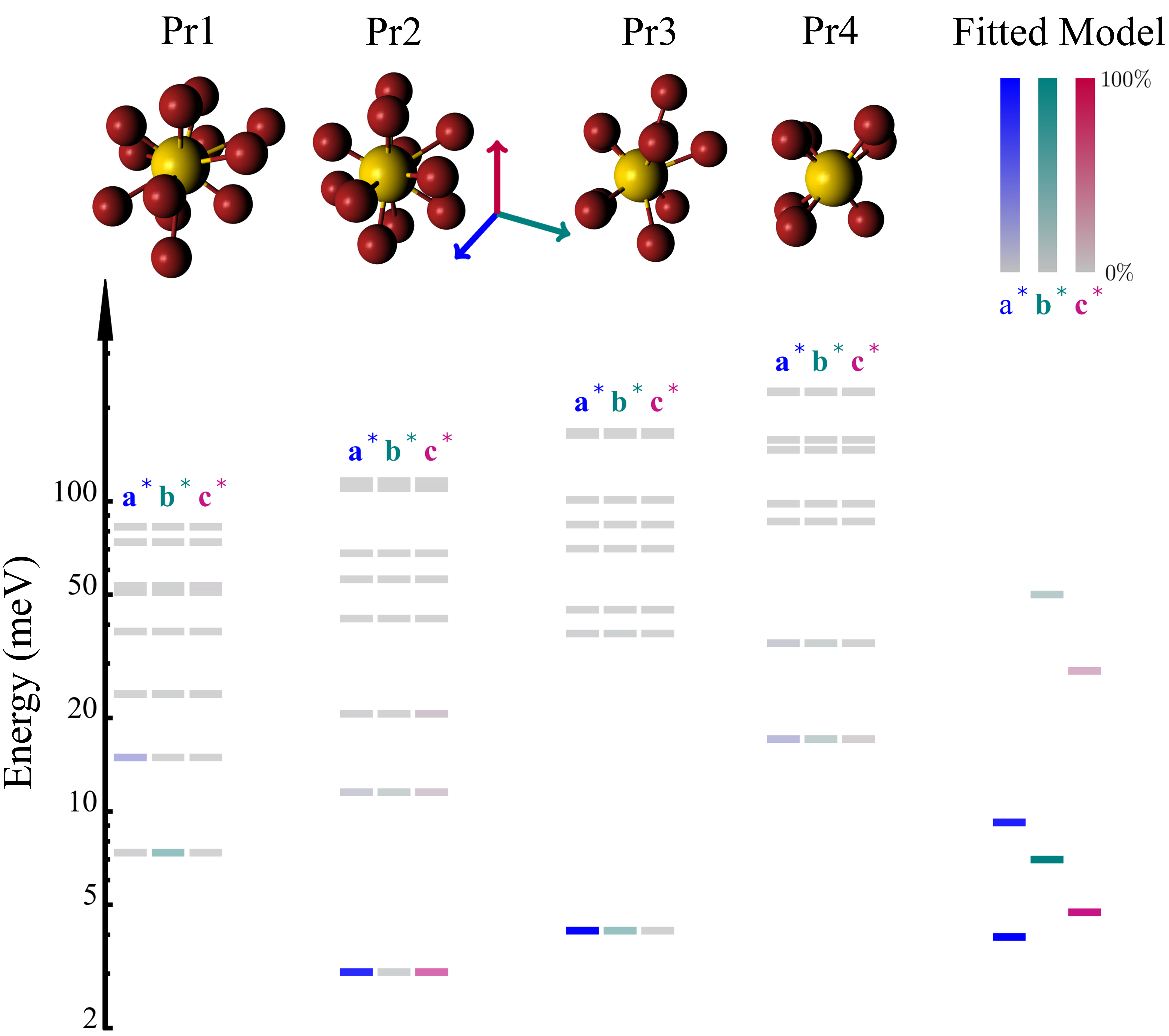}
\caption{Crystal-field level scheme calculated for each of the four distinct Pr$^{3+}$ Wyckoff sites in $\rm Pr_2Ti_2O_7$ based on the point charge model with a comparison to the site-averaged-level scheme inferred from susceptibility data (far right). The strength of each crystal-field level's van Vleck contribution to the magnetic susceptibility in the low-$T$ limit is indicated by the color of that level referring to the upper right color bars with cyan, green, and red associated with the three directions, respectively. ${\bf a^*}$, ${\bf b^*}$, and ${\bf c^*}$ are the reciprocal lattice directions normal to the (100), (010), and (001) planes. Gray levels do not contribute to the magnetic susceptibility at low $T$ for the indicated field direction. The coordinating O$^{2-}$ ligands that generate the crystal electric field are depicted as red spheres surrounding each of the four different Pr$^{3+}$ ions shown as yellow spheres.}
\label{Energystates}
\end{figure}

The high quality $\rm Pr_2Ti_2O_7$ single crystals described in the previous sections enabled us to explore the anisotropic magnetism of interacting non-Kramers rare-earth ions occupying four distinct low-symmetry sites. Prior experiments characterizing the physical properties of $\rm Pr_2Ti_2O_7$ are quite limited. Room-temperature ferromagnetism was reported for polycrystalline $\rm Pr_2Ti_2O_7$ nanoparticles grown through the sol-gel method. This effect is, however, unrelated to the rare-earth magnetism and appears to be connected with oxygen vacancies at the surfaces of nanoparticles \cite{PTO_nano1}. A Raman study on a single crystal of $\rm Pr_2Ti_2O_7$ synthesized by the float-zone method revealed structural stability up to 18 GPa and a complex spectrum as the low-symmetry monoclinic space group leads to 129 Raman active optical phonons \cite{PTO_Raman,PTOPSO_Raman}.

We measured the temperature-dependent magnetic susceptibility for both the single- and the polycrystalline Pr$_{2}$Ti$_{2}$O$_{7}$ samples. Figure \ref{SUSCfit}(a) shows $M(T)/B$ as a function of temperature for applied fields normal to the (001) plane ranging from 100 Oe to 70 kOe. No anomalies or hysteresis is observed down to $T=1.8$~K. This indicates Pr$_{2}$Ti$_{2}$O$_{7}$ has no magnetic phase transition in these temperature and field ranges. The field dependence of $M(T)/B$ at low $T$ for $B<15$~kOe indicates the presence of paramagnetic impurities. We use measurements of $M(T)$ at $B=1$~kOe to report the magnetic susceptibility $\chi(T)\equiv M(T)/B$ versus temperature and field direction in Fig. \ref{SUSCfit}(b). As should  be anticipated for the non-Kramers ion Pr$^{3+}$ in a monoclinic structure, there is considerable single-ion crystal electric-field-driven anisotropy with more than a factor 2 difference in low-$T$ susceptibility for the easy (100) versus the hard (001) directions. No field direction has indications of a magnetic phase transition. The inset shows the ratio of the single crystal to the powder sample magnetic susceptibility versus $T$, which provides a dimensionless measure of the magnetic anisotropy. The anisotropy is reduced at higher $T$, which indicates the population of excited crystal-field levels in that temperature range. An upturn in $\chi(T)$ is apparent at the lowest temperatures for both the single-crystal and polycrystalline samples. As detailed later, we associate this ``Curie tail'' with paramagnetic Pr$^{4+}$ impurities.

\section{Analysis and Discussion}

\subsection{Phenomenological crystal-field level scheme}
With the $J=4$ non-Kramers Pr$^{3+}$ ion in four different sites with $C_2^2$ symmetry there should be 36 crystal-field levels in $\rm Pr_2Ti_2O_7$. Neutron and/or Raman scatterings could, in principle, be used to determine these, but that is beyond the scope of this paper. By analyzing the $T-$dependence of the magnetic susceptibility, we can, however, obtain an estimate for the energy range associated with these crystal-field levels. For this purpose, we fit the data to the following three component phenomenological forms:
\begin{equation}
    \chi=\frac{C_0}{T}+2k_B\sum_i \frac{C_i}{\Delta_i}\rm tanh \frac{\beta \Delta_i}{2} .
    \label{hyper}
\end{equation}
Here, $\beta=1/k_BT$, and $k_B$ is the Boltzmann constant. The first term is the Curie term, which we will associate with Pr$^{4+}$impurities. The summations are over each of the four Pr$^{3+}$ sites, which are approximated as two-level systems with a singlet ground state and excited levels at a characteristic energy $\Delta_i$. This form is obtained as a low-$T$ approximation to the general Van Vleck susceptibility in the Supplementary Material Eq.(7) \cite{SM}. The expression has been constructed so that the high $T$ limit takes the Curie-form $C_i/T$ as for the impurity term. We found that two terms are sufficient to obtain excellent fits for all field directions.

The best fit parameters are listed in Table \ref{fitpam}. Consider first the fit to the poly-crystalline data. The sum of Curie constants $C_1+C_2=1.63(2)$~(emu$\cdot$K/Oe$\cdot$mole) agrees well with the Curie constant for  Pr$^{3+}$ $C_{\rm Pr^{3+}}=\mu_0N_Ag^2J(J+1)\mu_B^2/(3k_B)=1.60$~(emu$\cdot$K/Oe$\cdot$mole). The two polycrystal gap values provide a scale for the crystal-field levels from 5 to 28 meV. For comparison the crystal field levels for $\rm Pr_2Zr_2O_7$ are at [0, 10, 57, 82, 93, 109]~meV\cite{PZONC}. The ratio of the Curie constants $C_1/C_2=0.77(1)$ indicates similar spectral weight at $\Delta_1$ and $\Delta_2$. The ratio $C_0/(C_1+C_2)=0.53(1)$\% provides the order of magnitude of the Kramers rare-earth impurity concentration.

The fits to single-crystal data yield a broader range of gap values with the smallest gap value $\Delta_1=3.93(7)$~meV associated with the easy (100) direction and the largest gap value $\Delta_2=50(1)$~meV for the (010) direction. The different energies $\Delta_i^\alpha$ obtained for different field directions indicate the energy range for crystal-field levels $i$ with dominant contributions to the susceptibility for each field direction $\alpha$. However, with a total of 32 crystal-field transitions available from the ground state of $\rm Pr_2Ti_2O_7$, these data can only provide a general sense of the energy scale for the crystal-field levels that contribute most to the Van Vleck susceptibility for each field direction. The Curie constants $C_{i}^{\alpha\alpha}$ for each field direction $\alpha$ reflect the site averaged dipolar matrix elements for crystal-field levels separated by energies $\Delta^\alpha_i$ for $i>0$. The consistency between the directionally averaged single-crystal parameters and those obtained for polycrystalline samples supports the evidence from x-ray diffraction that the single crystal and powder samples are similar at the atomic scale despite the very different growth conditions. We note the Curie constant associated with impurities  $C_{0}^{\alpha\alpha}$ is more isotropic than for the intrinsic bulk terms $C_{1,2}^{\alpha\alpha}$. The impurity content as measured by $C_0$ is 40\% less in the single crystal as compared to the powder samples.

In this insulating rare-earth compound with shared ligands, super exchange interactions between praseodymium are expected to be on the 0.5-meV scale. This is an order of magnitude smaller than needed to close the lowest crystal-field energy gap. The low-symmetry environment, thus, effectively renders the non-Kramers rare-earth ion compound non-magnetic and precludes a phase transition or indeed any truly collective physics. It might, however, be possible to induce a quantum phase transition through the application of pressure as in TlCuCl$_3$\cite{TlCuCl3_PRL,TlCuCl3_JPSJ} and Tb$_2$Ti$_2$O$_7$\cite{TTO_pressure} before inducing a structural phase transition.

\subsection{Point charge model}
The splitting of the $J-$multiplet and the associated magnetic single-ion anisotropy is due to the crystal electric field (CEF) associated with the ligands surrounding each rare-earth ion. As detailed in the Supplementary Material\cite{SM}, the corresponding single ion Hamiltonian can be expressed in the form
\begin{equation}
\label{CEF}
   \mathcal{H}_{CEF}=\sum_{n,m} B_{n}^{m} O_{n}^{m},
\end{equation}
where O$_{n}^{m}$ are Stevens operators\cite{Hutchings,Stevens} and $B_n^m$ parametrize the effects of the CEF on the $J-$multiplet. Starting from the crystal structure, the point charge model yields estimates for $B_n^m$ which are provided for each distinct Pr-site within $\rm Pr_2Ti_2O_7$ in Table IV of the Supplementary Material\cite{SM}.

Diagonalizing Eq.~(\ref{CEF}) yields the CEF level scheme shown for each of the four Pr Wyckoff sites in Fig.~\ref{Energystates}.  The  predicted contribution of crystal field level to the low-$T$ van Vleck susceptibility for each of three field directions is listed in Table 5 of the Supplementary Material\cite{SM} and indicated by the color scheme in Fig.~\ref{Energystates}. The point charge model predicts that the Pr2 and Pr3 sites, which are ten fold coordinated by oxygen and bracket the perovskite (001) slabs (Fig.~\ref{structure}), are the dominant contributors to the low-$T$ van Vleck susceptibility (see also Fig. S1 of the Supplementary Material\cite{SM}).

The point charge model also indicates that the first excited CEF levels provide the largest contribution to the magnetic susceptibility. This support our use of the phenomenological form in Eq.~(\ref{hyper}) to fit the magnetic susceptibility data. Comparison of the point charge energy level scheme with that inferred from these fits shows a remarkable agreement in identifying the CEF levels that dominate the low-$T$ magnetic susceptibility for each of the three field directions. There is however, a consistent trend that the point charge model predicts lower-energy levels than inferred from the phenomenological fitting. This is also apparent in Fig.~\ref{SUSCfit}(b) where the point charge model (dashed lines) generally overestimates the low-$T$ magnetic susceptibility. Overall, although considering that there are no adjustable parameters, the point charge model does fairly well. For example, it correctly predicts that $\chi^{a^*}>\chi^{b^*}>\chi^{c^*}$. Our analysis shows that Eq.~(\ref{hyper}) provides an excellent account of the anisotropic $T$-dependent magnetic susceptibility of this singlet ground state system. The corresponding energy levels on the far right in Fig.\ref{Energystates} represent our best experimental estimate of the crystal-field level scheme in $\rm Pr_2Ti_2O_7$.

\subsection{Curie tails and Pr$^{4+}$ impurities}
Examining the low field behavior in greater detail, Fig. \ref{SUSCfit}(c) shows magnetization curves for various field directions at $T=5$~K. For all orientations, the data look linear on this scale. However, if we subtract a linear fit to the high-field regime from 6.5 to 7~T from the data, we obtain the non-linear component shown for the (001) direction in Fig. \ref{SUSCfit}(d). Resembling the magnetization curve for a paramagnetic impurity, the nonlinear component becomes more prominent when the thermal energy scale falls below the Zeeman energy scale. Under the assumption of a single characteristic impurity species it is possible to extract both the saturation magnetization and the impurity concentration from such data. To do so we fit the Langevin magnetization curve to the data,

\begin{equation}
   \Delta M(B,T)=fN_A\mu_{sat}L(\beta \mu_{sat}B).
   \label{langevin}
\end{equation}
Here $f$ is the paramagnetic impurity fraction, $N_A$ is Avagadro's number, and the Langevin function is given by  $L(x)=\lim_{J\rightarrow\infty}B_J(x)=\coth(x)-1/x$. $B_J(x)$ is the Brillouin function describing the magnetization curve for a paramagnetic impurity with spin-orbital angular momentum quantum number $J$. Figure \ref{SUSCfit}(d) shows that this functional form provides an excellent fit to the data with an impurity fraction of $f=0.96(2)\%$ and a saturation moment of $\mu_{sat}=2.00(3)~\mu_B$. The Curie constant corresponding to these parameters is approximately $f\mu_0N_A\mu_{sat}^2/(3k_B)=0.0048(2)$~(emu$\cdot$K/Oe$\cdot$mole), which is entirely consistent with $C_0$ in Table~\ref{fitpam} derived from $T$-dependent susceptibility data. Note that,  although the majority component of the magnetization is excluded from this analysis, the Langevin fit is consistent with the concentration estimate obtained from the ratio of Curie constants. We consider the concentration from the Langevin analysis to be more accurate, though, because it does not rely on the impurities having the same effective moment as the majority phase.

The specification of the $\rm Pr_6O_{11}$ starting material was $>99$\% rare-earth oxide with 99.99\% of the rare earth oxide being $\rm Pr_6O_{11}$. A non-Pr source for the paramagnetic impurity is, thus, unlikely. A possible explanation is that the rare-earth impurity is Pr$^{4+}$, which has a single $4f$ electron and the same magnetic properties as Ce$^{3+}$. In particular, it is a Kramers ion with a  saturation moment of  2.14 $\mu_B$, consistent with the Langevin analysis of the magnetization data.

\section{Conclusion}

We have successfully synthesized both polycrystalline samples and stoichiometric single crystals of Pr$_2$Ti$_2$O$_7$. Small levels of compositional deviations from the stoichiometric Pr$_2$Ti$_2$O$_7$ target compound lead to secondary phases of different, albeit related, structures. In contrast to the pyrochlore titanates, site disordering was not observed in the lower-symmetry monoclinic Pr$_2$Ti$_2$O$_7$ structure. 

Building upon the chemical and structural stabilities of monoclinic $\rm Pr_2Ti_2O_7$, we provide a process to grow large high-quality single crystals. Stoichiometric Pr$_2$Ti$_2$O$_7$ single crystals were grown using the TSFZ method. The susceptibility and magnetization measurements show no indications of a phase transition down to $T=1.8$~K. Our analysis of the susceptibility data indicates a singlet ground state with excited CEF levels at energies ranging from 3.93(7) to 50(1) meV. This is as anticipated for the low-symmetry crystal electric-field environment associated with Pr$^{3+}$ in monoclinic $\rm Pr_2Ti_2O_7$. At sufficiently low density, Kramers rare-earth impurities within $\rm Pr_2Ti_2O_7$ behave as isolated paramagnetic impurities that dominate over the majority phase paramagnetic van Vleck susceptibility at the lowest temperatures and fields ($T<5$~K, $B<5$~kOe). Analysis of the magnetization data in this regime yields a 0.96(2)\% impurity content with a saturation moment of $\mu_{sat}=2.00(3)~\mu_B$, which is consistent with a low concentration of Pr$^{4+}$ impurities.

The monoclinic structure contains a 3D network of Pr$^{3+}$  with edge-sharing ligand polyhedra assuring superexchange interactions on the 0.5 meV energy scale. If the gap between the two lowest-lying singlets can be closed through the application of hydrostatic of uniaxial pressure, or fields, a quantum phase transition into an ordered magnetic state induced by these interactions can be anticipated. $\rm Pr_2Ti_2O_7$, thus, may provide an interesting opportunity to explore universal aspects of quantum critical spin dynamics in a material where large high-quality crystals are attainable.

\section{Acknowledgements}
This work was supported as part of the Institute for Quantum Matter, an Energy Frontier Research Center funded by the U.S. Department of Energy, Office of Science, Basic Energy Sciences under Award No. DE-SC0019331. H.M., A.G. and C.L.B. were also supported by the Gordon and Betty Moore Foundation under Grant No. GBMF9456. The work at University of Houston (M.A. and C.W.C.) was supported by U. S. Air Force Office of Scientific Research Grants FA9550-15-1-0236 and FA9550-20-1-0068, the T. L. L. Temple Foundation, the John J. and Rebecca Moores Endowment, and the State of Texas through the Texas Center for Superconductivity at the University of Houston. Y.L. and C.L.C. were supported by U.S. Department of Energy, Basic Energy Science under Award Grant No. DESC0009390. M.L.T. and E.A. acknowledge funding, in part, from the U.S. Department of Energy, Office of Basic Energy Sciences through Contract No. DE-SC0020314. M.L.T. also acknowledges funding, in part, from the Office of Naval Research Multidisciplinary University Research Initiative (MURI) program through Contract No. N00014-20-1-2368.


\bibliography{ref}

\end{document}


\title{Supplementary Material\\
Quantum paramagnetism in a non-Kramers rare-earth oxide: Monoclinic $\rm Pr_2Ti_2O_7$}

\author{Huiyuan Man}
\affiliation{Institute for Quantum Matter and Department of Physics and Astronomy, The Johns Hopkins University, Baltimore, MD 21218, USA}

\author{Alireza Ghasemi}
\affiliation{Institute for Quantum Matter and Department of Physics and Astronomy, The Johns Hopkins University, Baltimore, MD 21218, USA}

\author{Moein Adnani}
\affiliation{Texas Center for Superconductivity and Department of Physics, University of Houston, Houston, TX 77204, USA}

\author{Maxime A. Siegler}
\affiliation{Department of Chemistry, The Johns Hopkins University, Baltimore, MD 21218, USA}

\author{Elaf A. Anber}
\affiliation{Department of Materials Science and Engineering, The Johns Hopkins University, Baltimore, MD 21218, USA}

\author{Yufan Li}
\affiliation{Department of Physics and Astronomy, The Johns Hopkins University, Baltimore, MD 21218, USA}

\author{Chia-Ling Chien}
\affiliation{Department of Physics and Astronomy, The Johns Hopkins University, Baltimore, MD 21218, USA}
\affiliation{Institute of Physics, Academia Sinica, Taipei 11519, Taiwan}
\affiliation{Department of Physics, National Taiwan University, Taipei 10617, Taiwan}

\author{Mitra L. Taheri}
\affiliation{Department of Materials Science and Engineering, The Johns Hopkins University, Baltimore, MD 21218, USA}

\author{Ching-Wu Chu}
\affiliation{Texas Center for Superconductivity and Department of Physics, University of Houston, Houston, TX 77204, USA}
\affiliation{Lawrence Berkeley National Laboratory, Berkeley, CA 94720, USA}

\author{Collin L. Broholm}
\affiliation{Institute for Quantum Matter and Department of Physics and Astronomy, The Johns Hopkins University, Baltimore, MD 21218, USA}
\affiliation{Department of Materials Science and Engineering, The Johns Hopkins University, Baltimore, MD 21218, USA}

\author{Seyed M. Koohpayeh}
\affiliation{Institute for Quantum Matter and Department of Physics and Astronomy, The Johns Hopkins University, Baltimore, MD 21218, USA}
\affiliation{Department of Materials Science and Engineering, The Johns Hopkins University, Baltimore, MD 21218, USA}
\affiliation{Ralph O'Connor Sustainable Energy Institute, Johns Hopkins University, Baltimore, MD 21218, USA}

\maketitle

\renewcommand{\thetable}{S\arabic{table}}
\renewcommand{\thefigure}{S\arabic{figure}}

\section{Solid State Synthesis of Polycrystalline Samples}

To explore the phase stability of Pr$_2$Ti$_2$O$_7$ we carried out solid state synthesis studies for a range of reagent compositions corresponding to Pr$_2$Ti$_{2+x}$O$_7$ with ($-0.16\leq x\leq0.16$). Powder x-ray diffraction confirmed the monoclinic structure with consistent lattice parameters for the Pr$_2$Ti$_2$O$_7$ majority phase in all of the samples. The phases observed, together with the lattice parameters of the main product (Pr$_2$Ti$_2$O$_7$), are reported in Table \ref{phase}. While the stoichiometrically synthesized powder seemed to be single-phase Pr$_2$Ti$_2$O$_7$ (within the detection limit of our x-ray diffraction measurements), the off-stoichiometric powders showed evidence of secondary phases of Pr$_{0.66}$TiO$_3$ and Pr$_4$Ti$_9$O$_{24}$ (for the Ti-rich samples) and Pr$_2$TiO$_5$ (for the Pr-rich samples). All of the prepared powder samples appeared green in color (as expected for Pr$^{3+}$)\cite{PZO_seyed} with slightly different shades.

Although no composition-temperature phase diagram is reported for the Pr$_2$O$_3$-TiO$_2$ system, the existence of the observed phases is in close agreement with the La$_2$O$_3$-TiO$_2$ phase diagram \cite{LTO_PhaseDiagram1,LTO_PhaseDiagram2},  the previously reported RE$_{2/3}$TiO$_3$ perovskite phase  \cite{La_Ti_O,La2_3TiO3,La2_3TiO3_x,Ln2_3TiO3,Pr6O11_TiO2}, and the orthorhombic phases of RE$_4$Ti$_9$O$_{24}$ and RE$_2$TiO$_5$ (RE = La and Pr) \cite{Pr6O11_TiO2,Ln2TiO5}.

The observation of secondary phases for the off-stoichiometric powders and the consistency of the measured lattice parameters for the  majority phase in all of the powders indicate that the structural stability of the low-symmetry monoclinic Pr$_2$Ti$_2$O$_7$ is significantly higher than that of cubic pyrochlore titanates, which are subject to disordering and non-stoichiometry \cite{HTO_Ali,ETO_Wang,YTO_Seyed}. To examine this further, the Ti-deficient powder (Pr$_2$Ti$_{2+x}$O$_7$, $x = -0.16$) was additionally sintered at a higher temperature in the halogen furnace with 53\% lamp power ($\approx 1500~^{\circ}{\rm C}-1600~^{\circ}$C) under ultra-high purity (UHP) argon gas. Although thermodynamic conditions to promote stuffing defects were in place in the form of Ti deficiency and synthesis at a higher temperature in a reducing atmosphere, the XRD results taken from this powder still showed two distinct phases of Pr$_2$Ti$_2$O$_7$ and Pr$_2$TiO$_5$ (as reported in Table \ref{phase}), rather than a stuffed-defective structure of Pr$_2$(Ti$_{2-x}$Pr$_x$)O$_{7-x}$. This  confirms that the monoclinic  Pr$_2$Ti$_2$O$_7$ structure does not support cation migration (the flexibility to move Pr$^{3+}$ into the Ti$^{4+}$ vacant sites). We hypothesize that this is due to the lower symmetry crystal structure and the large ratio of cation radii $r_{\rm Pr^{3+}} / r_{\rm Ti^{4+}}\approx$1.87. This is also consistent with an important contribution of the BO$_6$ octahedra (here B=Ti$^{4+}$) to the stability of the perovskite structure causing B ion vacancies to be rare, whereas A ion vacancies are common \cite{AdMat}. The stability of this structure is also confirmed by high-pressure Raman scattering studies \cite{PTO_RamanXray,PTO_Raman} and a report that La$_2$Ti$_2$O$_7$ does not undergo a structural phase transition until an external pressure of at least 16.7 GPa is applied\cite{LTO_HighPressure}.

For the off-stoichiometric powders, the extra Ti cations led to the formation of the secondary phase Pr$_{2/3}$TiO$_3$. This compound has a perovskite-derived structure with vacancies on one third of the cuboctahedral Pr$^{3+}$ sites due to the higher valence of the octahedral Ti$^{4+}$ cations \cite{Superionics}. On the other hand, for the Ti-deficient powders, the addition of Pr cations resulted in the orthorhombic secondary phase Pr$_2$TiO$_5$ (a Pr-rich phase compared to the Pr$_2$Ti$_2$O$_7$ majority phase) with the coexistence of 7 and 5 coordinated Pr$^{3+}$ and Ti$^{4+}$ cations, respectively \cite{Lanthanide_Titanates,Ln2TiO5,Ln2TiO5_CrystalChemistry}. These results show that small levels of compositional deviations (from the stoichiometric Pr$_2$Ti$_2$O$_7$ target compound) lead to secondary phases of different related structures.

With respect to the stoichiometric powder Pr$_2$Ti$_2$O$_7$, an additional higher-temperature synthesis (at $\approx 1500~^{\circ}{\rm C}-1600~^{\circ}$C, inside the image furnace under UHP argon gas) resulted in a lighter (neon) green colored powder and a better Rietveld refinement with a smaller $\chi^2 = 3.296$. The lattice constants were, however, measured to be the same, $a = 7.7116$ (\AA), $b = 5.4866$ (\AA), $c = 13.0041$ (\AA), and $\beta = 98.5426^{\circ}$. These measured values for the stoichiometric single-phase Pr$_2$Ti$_2$O$_7$ are comparably placed between the lattice parameters reported for the compounds La$_2$Ti$_2$O$_7$ and Nd$_2$Ti$_2$O$_7$ \cite{Lanthanide_Titanates}, and are consistent with the literature \cite{PTO_RamanXray,PTOstructure1,PTOstructure2,PTO_electronicstructure}. It is also worth noting that the high-temperature synthesis in argon did not show indications of oxygen (anion) vacancies or reduction of Ti$^{4+}$ into the darker colored Ti$^{3+}$. This is consistent with the limited accommodation of anion vacancies and poor reducibility of Ti in RE$_2$Ti$_2$O$_7$ (RE = La, Pr, Nd) layered perovskites processed in highly reducing atmospheres \cite{Ln2Ti2O7_Reduction}. In this recent study, a darker colored sample and limited loss of crystallinity were observed upon topochemical reductions of La$_2$Ti$_2$O$_7$ (i.e. by mixing with solid CaH$_2$, or flowing H$_2$), while a partial decomposition into La$_5$Ti$^{3.8+}_5$O$_{17}$ and La$_4$Ti$_3$O$_{12}$ was detected under more severe reducing conditions \cite{Ln2Ti2O7_Reduction}. These results again indicate the high structural stability of monoclinic Pr$_2$Ti$_2$O$_7$ compared with that of the pyrochlore titanates, in which anion vacancies and formation of Ti$^{3+}$ facilitate stuffing disorders and phase separation even in oxidizing atmospheres (of air and oxygen) during high-temperature synthesis of both powders \cite{ETO_Wang} and  float-zone grown single crystals \cite{HTO_Ali,ETO_Wang,YTO_Seyed}. It's important to bear this new information in mind when studying the monoclinic RE$_2$Ti$_2$O$_7$ materials.

To examine structural changes in an oxidizing atmosphere, the stoichiometric powder was additionally annealed at 1300 $^{\circ}$C in oxygen. This changed the color of the powder sample from a fresh green to a dark/brownish-green color due to the formation of Pr$^{4+}$ inclusions (a defect arising from the variable valence states of Pr), while no other apparent structural changes were observed within the detection limit of our powder x-ray diffraction measurements. In this regard, some levels of local structural distortions cannot be ruled out due to the structural environment around  Pr$^{4+}$ inclusions. Oxidation of Pr$^{3+}$ to Pr$^{4+}$ was observed at higher temperatures in an oxygen atmosphere near the melting point of Pr$_2$Ti$_2$O$_7$, which was accompanied by the formation of Ti-rich secondary phases, including Pr$_4$Ti$_9$O$_{24}$.

\begin{table*}
\normalsize
\caption{Observed phases and measured lattice parameters for polycrystalline samples of nominal compositions Pr$_{2}$Ti$_{2+x}$O$_{7}$ ($-0.16\leq x\leq0.16$) synthesized at 1300 $^{\circ}$C in air.}
\begin{center}
\renewcommand{\multirowsetup}{\centering}
{
\begin{tabular}{|c|c|c|c c c c|c|}
\hline
  $x$ & Main Phases\hspace{0.5cm} & Secondary Phase(s) & $a$ (\AA) & $b$ (\AA) & $c$ (\AA) & $\beta$ ($^{\circ}$) & $\chi^2$\\
\hline
-0.16 & Pr$_{2}$Ti$_{2}$O$_{7}$ \hspace{0.5cm} & Pr$_{2}$TiO$_{5}$, $\approx$ 9.7(2)wt\% & 7.7133 & 5.4873 & 13.0067 & 98.556 & 4.58\\
-0.08 & Pr$_{2}$Ti$_{2}$O$_{7}$ \hspace{0.5cm} & Pr$_{2}$TiO$_{5}$, $\approx$ 7.7(6)wt\% & 7.7132 & 5.4874 & 13.0083 & 98.548 & 2.52\\
0 & Pr$_{2}$Ti$_{2}$O$_{7}$ \hspace{0.5cm} & $-$ & 7.7118 & 5.4863 & 13.0041 & 98.552 & 2.83\\
0.08 & Pr$_{2}$Ti$_{2}$O$_{7}$ \hspace{0.5cm} & Pr$_{0.66}$TiO$_{3}$, $\approx$ 3.2(4)wt\% & 7.7134 & 5.4875 & 13.0088 & 98.554 & 3.67\\
0.16 & Pr$_{2}$Ti$_{2}$O$_{7}$ \hspace{0.5cm} & Pr$_{0.66}$TiO$_{3}$, $\approx$ 4.2(1)wt\%, Pr$_{4}$Ti$_{9}$O$_{24}$, $\approx$ 0.9(9)wt\% & 7.7139 & 5.4879 & 13.0098 & 98.553 & 2.22\\
\hline
\end{tabular}
}
\end{center}
\label{phase}
\end{table*}

\section{Single Crystal X-ray Diffraction}

The structure of our $\rm Pr_2Ti_2O_7$ single crystals was determined using single crystal X-ray diffraction and the full results are presented here. The  data were acquired using a SuperNova diffractometer (equipped with Atlas detector) with Mo $K_\alpha$ radiation ($\lambda$ = 0.71073 Å) under the program CrysAlisPro (Version CrysAlisPro 1.171.39.29c\cite{Rigaku}). The same program was used to refine the cell dimensions and for data reduction. The structure was solved with the program SHELXS-2018/3 \cite{SHELXSHELXL} and was refined on F$^{2}$ with SHELXL-2018/3 \cite{SHELXSHELXL}. Empirical absorption corrections using spherical harmonics was applied using CrysAlisPro. The data were acquired with the crystal cooled to 213(2) K by a  Cryojet system from Oxford Instruments. The occupancy of all four Pr sites and all four Ti sites were freely refined  and are presented in Table \ref{occu}. The absolute configuration was established by anomalous dispersion effects in diffraction measurements on the crystal, and the Flack and Hooft parameters refine to -0.024(11) and -0.026(9), respectively.

\newpage 
\begin{table*}[h]
\normalsize
\caption{Single-crystal data and structure refinement\cite{Refinement_Max} for Pr$_2$Ti$_2$O$_7$ at 213(2) K.}
\begin{center}
\renewcommand{\multirowsetup}{\centering}
{
\begin{tabular}{|p{5cm}|p{8cm}| }

\multicolumn{2}{l}{}  \\

\multicolumn{2}{l}{\bf Crystal data} \\
\hline
Chemical formula & Pr$_{7.72}$Ti$_{7.70}$O$_{28}$ \\
\hline
$M_r$ & 1904.66 \\
\hline
Crystal system & Monoclinic \\
\hline
Space group & $P$2$_1$ \\
\hline
Temperature (K) & 213(2) \\
\hline
$a$, $b$, $c$ (\AA) & 7.72545(11), 5.49714(9), 13.03798(19) \\
\hline
$\alpha$, $\beta$, $\gamma$ ($^{\circ}$) & 90, 98.5651(13), 90 \\
\hline
$V$ (\AA$^3$) & 547.52 (1) \\
\hline 
$Z$ &  1\\
\hline
Radiation type & Mo $K_{\alpha}$  \\
\hline
$\mu$ (mm$^{-1}$) & 19.56 \\
\hline
Crystal size (mm) & $0.13 \times 0.12 \times 0.08$ \\
\hline
\multicolumn{2}{l}{}  \\

\multicolumn{2}{l}{\bf Data collection}  \\
\hline
Diffractometer & SuperNova, Dual, Cu at zero, Atlas \\
\hline
Absorption correction & Multi-scan,  
\textit{CrysAlisPRO} \cite{Rigaku} Empirical absorption correction using spherical harmonics, implemented in SCALE3 ABSPACK scaling algorithm.
 \\
\hline
 $T_{min}$, $T_{max}$ & 0.697, 1.000 \\
\hline
No. of measured, independent and observed [$I > 2\sigma(I)$] reflections & 20945, 4142, 4114 \\
\hline
$R_{int}$ & 0.039 \\
\hline
$(sin\theta/\lambda)_{max}$ (\AA$^{-1}$) & 0.766 \\
\hline
\multicolumn{2}{l}{}  \\

\multicolumn{2}{l}{\bf Refinement}  \\
\hline
$R[F^2 > 2\sigma(F^2)]$, $wR(F^2)$, $S$ & 0.017, 0.039, 1.12\\
\hline
No. of reflections & 4142 \\
\hline
No. of parameters & 208 \\
\hline
No. of restraints & 1 \\
\hline
$\Delta\rho_{max}$, $\Delta\rho_{min}$(e\AA$^{-3}$) &	1.26, -1.00  \\
\hline
Absolute structure & Flack x determined using 1850 quotients [(I+)-(I-)]/[(I+)+(I-)]\cite{Parson} \\
\hline
Absolute structure parameter & -0.024 (11)  \\
\hline
Computer programs &\textit{CrysAlisPRO} \cite{Rigaku},\textit{SHELXS2018/3} \cite{SHELXSHELXL}, \textit{SHELXL2018/3}\cite{SHELXSHELXL},\textit{SHELXTL} v6.10 \cite{HistorySHELX}. \\
\hline
\end{tabular}
}
\end{center}
\label{refine}
\end{table*}
\newpage

\begin{table*}[h]
\normalsize
\caption{ Atomic coordinates and isotropic displacement parameters $U$\textsubscript{eq}  for Pr$_2$Ti$_2$O$_7$ at 213(2) K with estimated standard deviations in parentheses.}
\begin{center}
\renewcommand{\multirowsetup}{\centering}
{
\begin{tabular}{|c|c|c|c|c|c|}
\hline
label & $x$ & $y$ & $z$ & occupancy & $U$\textsubscript{eq} (\AA\textsuperscript{2}) \\
\hline
Pr1 & 0.71986(3) & 0.74940(5) & 0.88289(2) & 0.970(5) & 0.00365(8) \\
\hline
Pr2 & 0.22763(3) & 0.74203(5) & 0.90517(2) & 0.969(5) & 0.00393(7) \\
\hline
Pr3 & 0.64906(3) & 0.21122(6) & 0.61231(2) & 0.959(5) & 0.00534(8) \\
\hline
Pr4 & 0.14587(3) & 0.15391(5) & 0.57704(2) & 0.961(5) & 0.00361(8) \\
\hline
Ti1 & 0.96851(10) & 0.2374(2) & 0.88149(6) & 0.966(6) & 0.0028(2) \\
\hline
Ti2 & 0.47236(10) & 0.2384(2) & 0.87972(6) & 0.964(6) & 0.0028(2) \\
\hline
Ti3 & 0.92390(11) & 0.71372(18) & 0.67876(6) & 0.963(6) & 0.0031(2) \\
\hline
Ti4 & 0.41587(11) & 0.70808(17) & 0.67475(6) & 0.955(6) & 0.0027(2) \\
\hline 
O1 & 0.2230(4) & 0.1930(7) & 0.8923(3) & 1 & 0.0080(8)\\
\hline
O2 & 0.7268(4) & 0.3190(7) & 0.9089(3) & 1 & 0.0062(7) \\
\hline
O3 & 0.9625(4) & 0.9732(7) & 0.9815(3) & 1 & 0.0067(7) \\
\hline
O4 & 0.5283(5) & 0.9732(7) & 0.9795(3) & 1 & 0.0076(7) \\
\hline
O5 & 0.8892(4) & 0.0404(7) & 0.7726(3)& 1 & 0.0070(7) \\
\hline
O6 & 0.4972(5) & 0.0462(7) & 0.7701(3) & 1 & 0.0069(7)\\
\hline
O7 & 0.9742(4) & 0.5524(7) & 0.8159(3) & 1 & 0.0062(7)\\
\hline
O8 & 0.4359(4) & 0.5561(7) & 0.8173(3) & 1 & 0.0056(7)\\
\hline
O9 & 0.9185(5) & 0.4478(7) & 0.5923(3) & 1 & 0.0067(7)\\
\hline
O10 & 0.3788(5) & 0.4289(7) & 0.6070(3) & 1 & 0.0080(7)\\
\hline
O11 & 0.8739(5) & 0.9341(7) & 0.5699(3) & 1 & 0.0058(7)\\
\hline
O12 & 0.4062(5) & 0.9202(7) & 0.5595(3) & 1 & 0.0071(7)\\
\hline
O13 & 0.6734(4) & 0.6573(8) & 0.6928(3) & 1 & 0.0077(7) \\
\hline
O14 & 0.1742(4) & 0.8015(7) & 0.6962(3) & 1 & 0.0078(7)\\
\hline
\end{tabular}
}
\end{center}
\label{occu}
\end{table*}

\section{The Point Charge Model and the Magnetic Susceptibility}

The crystal electric field (CEF) at the magnetic site plays an important role in rare earth magnetism. Here we provide a brief summary of the theory based on the papers by Hutchings\cite{Hutchings} and Scheie\cite{AllenCF} and its application to $\rm Pr_2Ti_2O_7$. The effect of the CEF on rare earth magnetism can be described through a single ion spin hamiltonian of the form
\begin{equation}
   \mathcal{H}_{CEF}=\sum_{n,m} B_{n}^{m} O_{n}^{m}.
\label{eq:hcef}
\end{equation}
Here O$_{n}^{m}$ are Stevens operators\cite{Stevens} and $B_n^m$, parametrize the effects of the CEF. With knowledge of the local coordinating environment, the point charge model yields the following estimates for $B_n^m$
\begin{equation}
    B_{n}^{m}=\sum_{j=1}^k\frac{1}{4\pi\epsilon_0}\frac{-|e|q_j}{R_j}\frac{\langle r^n\rangle}{R_j^n}D_{nm}(J)Z_{nm}(\hat{\bf R}_j).
\label{eq:bnm}
\end{equation}
Here the summation is over the $k$ ligands which are treated as point charges with charge $q_j$ displaced by  ${\bf R}_j$ from the rare earth ion. $\langle r^n\rangle$ indicate radial averages characterizing the spatial extent of the $4f$ electron wave function. $D_{nm}(J)$ consolidate dimensionless factors associated with the Taylor expansion of the ligand potential and application of the Wigner-Eckart theorem
\begin{equation}
D_{nm}(J)=\frac{4\pi}{2n+1}\theta_n(J)C_{nm}
\end{equation}

\begin{table*}[h]
\normalsize
\caption{CEF parameters in Eq.~\ref{eq:hcef} for the four different Pr$^{3+}$ sites in $\rm Pr_2Ti_2O_7$ as determined by the point charge model using Eq.~\ref{eq:bnm}\cite{AllenCF}.}
\begin{center}
\renewcommand{\multirowsetup}{\centering}
{
\begin{tabular}
{|c|c|c|c|c|}
\hline
$B_n^m$ & Pr$_1$ (meV) & Pr$_2$ (meV)  & Pr$_3$ (meV)  & Pr$_4$ (meV) \\
\hline
$B_{2}^{-2}$ & 0.1822 & 0.8109 & 0.9893 & -1.7942 \\
\hline
$B_{2}^{-1}$ & -0.2733 & 8.1666 & -5.0499 & 4.0759 \\
\hline
$B_{2}^{0}$ & 0.2795 & 0.2990 & 0.1232 & 4.6609 \\
\hline
$B_{2}^{1}$ & -0.6971 & 2.9085 & -3.1685 & -4.4919 \\
\hline
$B_{2}^{2}$ & -1.3554 & -1.175 & -5.1517 & -1.6264 \\
\hline
$B_{4}^{-4}$ & -0.0238 & -0.0185 & 0.0144 & -0.0258 \\
\hline
$B_{4}^{-3}$ & 0.2781 & 0.2808 & -0.0371 & 0.3174 \\
\hline 
$B_{4}^{-2}$ & 0.0299 & 0.0325 & 0.0138 & 0.0224 \\
\hline
$B_{4}^{-1}$ & 0.0058 & -0.011 & -0.1023 & -0.0362 \\
\hline
$B_{4}^{0}$ & 0.0036 & -0.0074 & -0.0011 & -0.0037 \\
\hline
$B_{4}^{1}$ & 0.0035 & 0.0126 & 0.0350 & 0.0586 \\
\hline
$B_{4}^{2}$ & 0.0247 & -0.0139 & 0.0435 & 0.1032 \\
\hline
$B_{4}^{3}$ & -0.0237 & 0.0015 & -0.0230 & -0.0489 \\
\hline
$B_{4}^{4}$ & -0.0024 & -0.0043 & 0.0497 & 0.1166 \\
\hline
$B_{6}^{-6}$ & 0.0005 & 0.0004 & -0.0002 & -0.0005 \\
\hline
$B_{6}^{-5}$ & -0.0050 & -0.0049 & 0.0015& 0.0062 \\
\hline
$B_{6}^{-4}$ &  -0.0010 & -0.0010 & 0.0002 & -0.0002 \\
\hline
$B_{6}^{-3}$ & 0.0022 & 0.0015 & 0.0014 & 0.0004 \\
\hline
$B_{6}^{-2}$ & 0.0009 & 0.0009 & 0.0001 & 0.0014 \\
\hline
$B_{6}^{-1}$ & -0.0005 & -0.0007 & -0.0004 & -0.0023 \\
\hline
$B_{6}^{0}$ & 0.0003 & 0.0003 & -0.0001 & -0.0003 \\
\hline
$B_{6}^{1}$ & -0.0032 & -0.0032 & 0.0012 & 0.0013 \\
\hline
$B_{6}^{2}$ & -0.0012 & -0.0013 & -0.0002 & -0.0004 \\
\hline
$B_{6}^{3}$ & -0.0011 & -0.0003 & 0.0006& 0.0022 \\
\hline
$B_{6}^{4}$ & -0.0034 & -0.0024 & 0.0005 & 0.0029 \\
\hline
$B_{6}^{5}$ & 0.0012 & 0.0009 & -0.0059 & -0.0071 \\
\hline
$B_{6}^{6}$ & -0.0045 & -0.0040 & -0.0037 & -0.0036 \\
\hline
\end{tabular}
}
\end{center}
\label{cflvls}
\end{table*}

Referring to Hutchings\cite{Hutchings}, $\theta_n(J)$ are given in table VI and $C_{nm}$ (consistent with the definition by Scheie\cite{AllenCF}) are the numerical pre-factors in front of the square brackets in table IV.
$Z_{nm}({\hat\bf r})$ are tesseral harmonic functions\cite{Hutchings,Stevens}. The crystal field parameters $B_n^m$ for $\rm Pr_2Ti_2O_7$ calculated from Eq.~\ref{eq:bnm} using $PyCrystalField$\cite{AllenCF} are listed in table~\ref{cflvls}.

Denoting the eigenstates and eigenvalues of Eq.~\ref{eq:hcef} by $|\gamma\rangle$ and $E_\gamma$ respectively, the magnetic susceptibility is obtained through linear response theory:

\begin{equation}
   \chi^{\alpha\alpha}(T)= n (g_J \mu_B)^2 \sum_{\gamma} \frac{\langle \gamma| J^{\alpha} | \gamma \rangle|^2}{k_BT} \rho_\gamma  + 
   n (g_J \mu_B)^2\sum_{\gamma' \neq \gamma}\frac{|\langle \gamma| J^{\alpha} |{\gamma'} \rangle|^2}{E_{\gamma'}-E_\gamma}(\rho_\gamma-\rho_{\gamma'}).
   \label{VVSusc}
\end{equation}
Here $n$ is the rare earth number density, $\rho_\gamma=\exp(-\beta E_\gamma)/{\cal Z}$ is the population factor for CEF level $\gamma$, $\beta=(k_BT)^{-1}$, $k_B$ is the Boltzmann constant, and $\mathcal{Z}$ is the partition function. The first Curie-term is associated with degenerate crystal field levels which do not occur for non-Kramers ions such as $\rm Pr^{3+}$ in a low symmetry environment as in $\rm Pr_2Ti_2O_7$. This term is however present for the $\rm Pr^{4+} $ Kramers ion. The second term describes the Van Vleck susceptibility, which is associated with transitions between distinct crystal field levels. The $T\rightarrow 0$ limit of the Van Vleck susceptibility is 
\begin{equation}
\lim_{T\rightarrow 0}\chi^{\alpha\alpha}_{vv}(T)=\sum_{\gamma\neq 0}\chi^{\alpha\alpha}_{\gamma},
\end{equation}
Where the low $T$ susceptibility associated with each CEF level is 
\begin{equation}
   \chi^{\alpha\alpha}_{\gamma}=n (g_J \mu_B)^2 \frac{2|\langle \gamma| J^{\alpha} |\gamma' \rangle|^2}{\Delta_{\gamma}}
   \label{eq:chig0}
\end{equation}
and we have introduced the excited state energy $\Delta_\gamma=E_\gamma-E_0$. When the lowest lying CEF level $\gamma=1$ is the dominant contribution to $\chi^{\alpha\alpha}_{vv}(T)$ we have 
\begin{equation}
    \chi^{\alpha\alpha}_{vv}(T)\approx \chi^{\alpha\alpha}_{1} \tanh\frac{\beta\Delta_1}{2}
\end{equation}
A superposition of terms of this form was found to provide an excellent fit to the susceptibility data for $\rm Pr_2Ti_2O_7$ (main text Fig. 5(b)). To determine the contribution of each CEF level to $\chi^{\alpha\alpha}_{vv}(T)$ within the point charge model $\chi^{\alpha\alpha}_{\gamma}$ was calculated for each of the four praseodymium sites in $\rm Pr_2Ti_2O_7$ using the $PyCrystalField$ software\cite{AllenCF} and is presented in Table S5. $\chi^{\alpha\alpha}_{\gamma}$ were also used to indicate the relative contributions of each crystal field level to the low $T$ magnetic susceptibility in Fig.6 of the main text.

\begin{table*}[h]
    \caption{Low $T$ Van Vleck susceptibility associated with each crystal field level of each Pr$^{3+}$ site in $\rm Pr_2Ti_2O_7$. Calculated using Eq.~\ref{eq:chig0} based on the point charge model with CEF parameters given in Table~\ref{cflvls}.}
    \begin{minipage}{.3\linewidth}
      \centering
        \begin{tabular}{|c|c|c|c|}
        \hline
        \multicolumn{4}{|c|}{Pr$_1$} \\
        \hline
        $\Delta_\gamma$ & $\chi_\gamma^{a^*}$ & $\chi_\gamma^{b^*}$ & $\chi_\gamma^{c^*}$  \\
        \hline
        meV&\multicolumn{3}{|c|}{emu/Oe/mole Pr} \\
        \hline
        7.35&0.0002&0.0090&0\\
        14.90&0.0053&0&0.0001\\
        23.88&0&0.0005&0.0002\\
        37.99&0&0&0\\
        50.88&0.0001&0.0001&0.0001\\
        53.32&0&0.0002&0.0003\\
        73.80&0&0&0\\
        82.79&0&0&0\\
        \hline
        $\sum_\gamma\chi^{\alpha\alpha}_\gamma$&0.0056&0.0098&0.0007\\
        \hline
        \end{tabular}
    \end{minipage}%
    \begin{minipage}{.3\linewidth}
      \centering
        \begin{tabular}{|c|c|c|c|}
        \hline
        \multicolumn{4}{|c|}{Pr$_2$} \\
        \hline
        $\Delta_\gamma$  & $\chi_\gamma^{a^*}$ & $\chi_\gamma^{b^*}$ & $\chi_\gamma^{c^*}$  \\
        \hline
        meV&\multicolumn{3}{|c|}{emu/Oe/mole Pr} \\
        \hline
        3.02&0.0224&0.0008&0.0161\\
        11.51&0.0012&0.0016&0.0021\\
        20.62&0.0002&0.0001&0.0001\\
        41.85&0&0.0001&0\\
        55.94&0&0&0\\
        67.88&0.0001&0&0.0001\\
        110.03&0&0.001&0\\
        116.18&0&0&0\\
        \hline
        $\sum_\gamma\chi^{\alpha\alpha}_\gamma$&0.0239&0.0026&0.0184\\
        \hline
        \end{tabular}
    \end{minipage}
    \begin{minipage}{.3\linewidth}
      \centering
        \begin{tabular}{|c|c|c|c|}
        \hline
        \multicolumn{4}{|c|}{Pr$_3$} \\
        \hline
        $\Delta_\gamma$  & $\chi_\gamma^{a^*}$ & $\chi_\gamma^{b^*}$ & $\chi_\gamma^{c^*}$  \\
        \hline
        meV&\multicolumn{3}{|c|}{emu/Oe/mole Pr} \\
        \hline
        4.12&0.0264&0.0087&0.0005\\
        37.40&0.0002&0.0008&0\\
        44.69&0&0&0.0002\\
        70.27&0.0001&0&0\\
        84.15&0&0&0\\
        100.94&0&0&0\\
        163.56&0&0&0\\
        167.52&0&0&0\\
        \hline
        $\sum_\gamma\chi^{\alpha\alpha}_\gamma$&0.0267&0.0095&0.0007\\
        \hline
        \end{tabular}
    \end{minipage}%
    \begin{minipage}{.3\linewidth}
      \centering
        \begin{tabular}{|c|c|c|c|}
        \hline
        \multicolumn{4}{|c|}{Pr$_4$} \\
        \hline
        $\Delta_\gamma$  & $\chi_\gamma^{a^*}$ & $\chi_\gamma^{b^*}$ & $\chi_\gamma^{c^*}$  \\
        \hline
        meV&\multicolumn{3}{|c|}{emu/Oe/mole Pr} \\
        \hline
        17.08&0.0036&0.0027&0.0006\\
        34.86&0.0011&0.0011&0.0001\\
        86.04&0&0&0\\
        98.07&0&0&0\\
        146.43&0&0&0\\
        157.80&0&0&0\\
        224.70&0&0&0\\
        225.98&0&0&0\\
        \hline
        $\sum_\gamma\chi^{\alpha\alpha}_\gamma$&0.0047&0.0038&0.0007\\
        \hline
        \end{tabular}
    \end{minipage}   
\end{table*}


In the main text the magnetic susceptibility calculated by the point charge model using  $PyCrystalField$\cite{AllenCF} was compared to the experimental data in Fig. 5(b). Considering that there are no adjustable parameter the fit is acceptable. In Fig. S1 we show the point charge model calculation of the contribution of each Pr site to the magnetic susceptibility measured along three directions. Pr2 and Pr3 contribute most to the susceptibility with some contributions of Pr1 for fields along $(010)$ and generally minimal contributions from Pr4, which has the highest lying CEF levels (see Fig. 6). 
\newpage

\begin{figure*}[h]
\centering
\includegraphics[width=15cm]{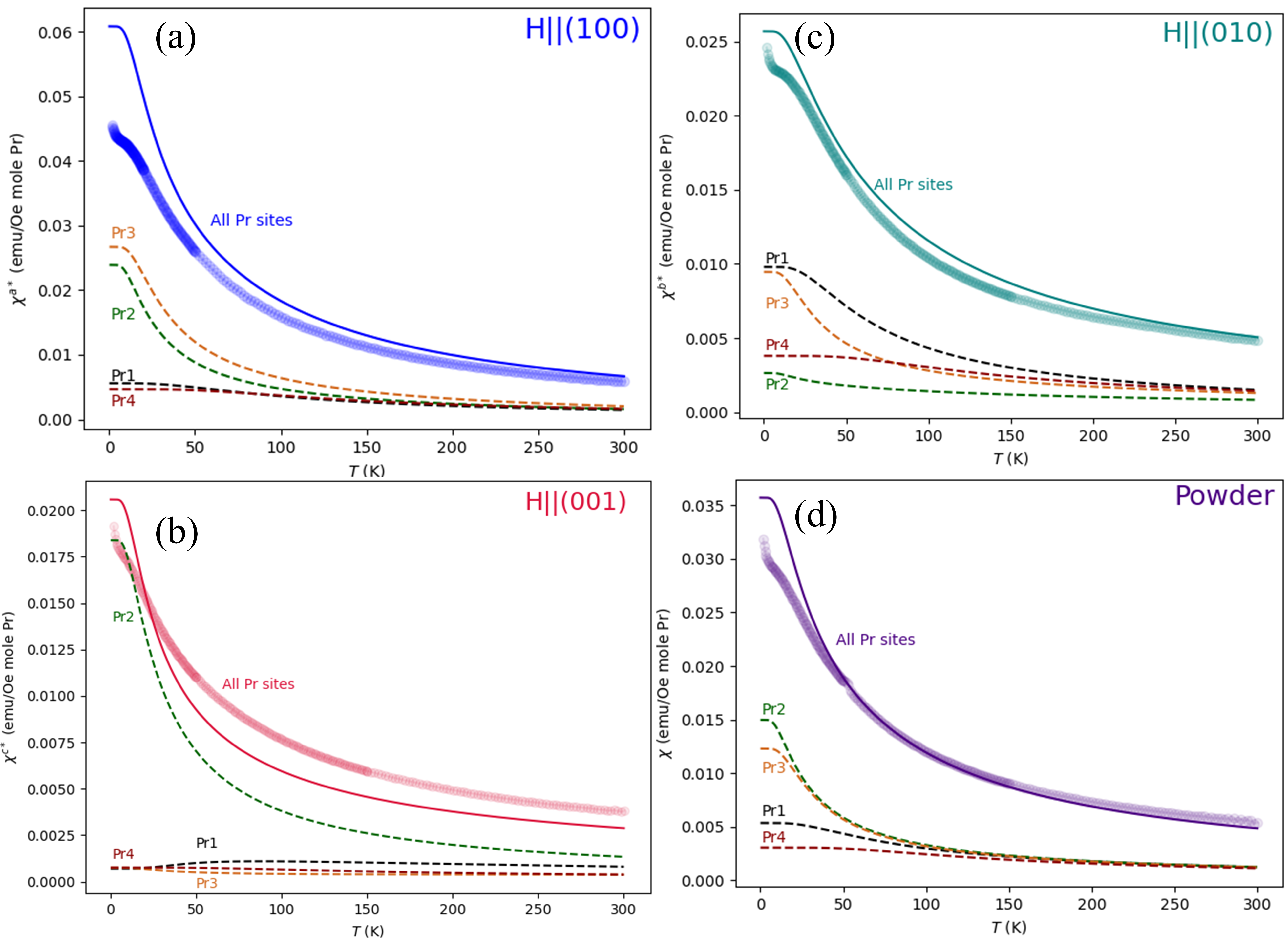}
\caption{Magnetic susceptibility of $\rm Pr_2Ti_2O_7$ measured for fields along (a) (100), (b) (001), (c) (010), and (d) for a powder sample. The data are compared to the predictions of the point charge model (Eq.~\ref{VVSusc})\cite{AllenCF} for the four distinct Pr$^{3+}$ Wyckkoff sites individually (dashed lines) and superimposed (solid line).}
\label{SUSCcalc}
\end{figure*}

\bibliography{ref}